\documentclass[prd,superscriptaddress,showpacs,floatfix,%
preprintnumbers, nofootinbib,hyperref]{revtex4} 
\usepackage{graphicx}
\usepackage{url}

\newcommand{\ben}{\begin{equation}}
\newcommand{\een}{\end{equation}}

\newcommand{\be}{\begin{equation}}
\newcommand{\ee}{\end{equation}}  
\newcommand{\bea}{\begin{eqnarray}}
\newcommand{\eea}{\end{eqnarray}}  
\newcommand{\gag}{g_{a\gamma}}

\begin{document}
\preprint{LMU-ASC 44/12, MPP-2012-108}

\title{Hardening of TeV gamma spectrum of active galactic nuclei in galaxy clusters
by conversions of photons into axion-like particles}

\author{Dieter Horns} 
\affiliation{Institut f\"ur Experimentalphysik, Universit\"at Hamburg,
Luruper Chaussee 149, 22761 Hamburg, Germany}
\author{Luca Maccione}
\affiliation{Arnold Sommerfeld Center, Ludwig-Maximilians-Universit\"at, Theresienstra{\ss}e 37, 80333 M\"unchen, Germany} 
\affiliation{Max-Planck-Institut f\"ur Physik (Werner-Heisenberg Institut), F\"ohringer Ring 6, 80805 M\"unchen, Germany}

\author{Manuel Meyer}
\affiliation{Institut f\"ur Experimentalphysik, Universit\"at Hamburg,
Luruper Chaussee 149, 22761 Hamburg, Germany}
\author{Alessandro Mirizzi}
\affiliation
{II Institut f\"ur Theoretische Physik, Universit\"at Hamburg, Luruper Chaussee 149, 22761 Hamburg, Germany} 
\author{Daniele Montanino}
 \affiliation{Dipartimento di Matematica e Fisica ``Ennio de Giorgi'', Universit\`a del Salento\
and Sezione INFN di Lecce,
Via Arnesano, 
I--73100 Lecce, Italy
}
 \author{Marco Roncadelli} 
\affiliation{INFN, Sezione di Pavia, Via A.~Bassi 6, 27100 Pavia,
Italy}

\begin{abstract}
A  fraction of AGN producing VHE $\gamma$-rays are located in galaxy clusters. The magnetic field
present in the intra-cluster medium would lead to conversions of VHE photons into axion-like particles (ALPs), 
which are a generic prediction of several extensions of the Standard Model. 
ALPs produced in this way would traverse cosmological distances unaffected by the extragalactic background light at variance with VHE photons which undergo a substantial absorption.
Eventually, a nontrivial  fraction of ALPs would re-convert into VHE photons in the magnetic field of the Milky Way. This mechanism  produces a significant hardening of the VHE spectrum of AGN in galaxy clusters. 
As a specific example we consider the energy spectra of two observed VHE $\gamma$-ray sources located in galaxy clusters, namely 1ES 0414+009 at redshift $z=0.287$ and Mkn 501 at $z=0.034$.  
We find that the hardening in the observed spectra becomes relevant at $E \gtrsim 1$~TeV.
The detection of this signature would allow to indirectly probe the existence of ultra-light ALPs with mass $m_a \lesssim 10^{-8}$~eV and  photon-ALP coupling $g_{a\gamma} \lesssim 10^{-10}$~GeV$^{-1}$
with the presently operating Imaging Atmospheric Cherenkov Telescopes like H.E.S.S., MAGIC, VERITAS and CANGAROO-III and even more likely with the planned detectors like CTA, HAWC and Hundred Square-km Cosmic ORigin Explorer (HiSCORE).
An independent laboratory check of ultra-light ALPs invoked in this mechanism can  be  performed  with the planned upgrade of the photon regeneration experiment Any Light Particle Search (ALPS) at Deutsches Elektronen-Synchrotron and with  the next generation solar axion detector International Axion Observatory.
\end{abstract}



\maketitle

\section{Introduction}

Axion-like particles (ALPs) are very light pseudo-scalar bosons $a$ with a two-photon coupling $a \gamma \gamma$ which 
are predicted by several extensions of the Standard Model like four-dimensional ordinary and supersymmetric models (see e.g.~\cite{susy1,susy4}), Kaluza-Klein theories (see e.g.~\cite{kaluzaklein1}) and especially superstring theories (see e.g.~\cite{witten1,witten2,witten3,witten4}) (for a review, see~\cite{masso,Jaeckel:2010ni}). In the presence of an external magnetic field, the $a \gamma \gamma$ coupling entails that interaction eigenstates differ from propagation eigenstates thereby leading to the phenomenon of photon-ALP conversion $\gamma \leftrightarrow a$ and in particular to photon-ALP oscillations~\cite{sikivie,Raffelt:1987im,Anselm:1987vj}. This mixing effect is exploited to search for generic ALPs in light-shining-through-the-wall experiments (see e.g. the ALPS~\cite{Ehret:2010mh} and the GammeV~\cite{gammev} experiments), for solar ALPs (see e.g. the CAST experiment~\cite{Arik:2011rx}) and for ALP dark matter~\cite{Arias:2012az,sikivie1} in micro-wave cavity experiments (see e.g. the ADMX experiment~\cite{Duffy:2006aa}).

Photon-ALP oscillations would also lead  to intriguing signatures in astrophysical and
cosmological observations~\cite{Jaeckel:2010ni,Mirizzi:2007hr,Dupays:2005xs}.
In particular, over the last few years it has been realized that the
$a\gamma\gamma$ coupling can also produce detectable effects in the
observations of distant active galactic nuclei (AGN), since photons emitted by
these sources can mix with ALPs during their propagation
through large-scale magnetic fields~\cite{De Angelis:2007yu}. In this context, photon-ALP
 oscillations~\cite{HS:2007,HSI:2007,De Angelis:2007yu} 
provide a natural mechanism to drastically reduce the 
absorption of  very high-energy
(VHE) photons   by pair production processes ($\gamma_{\rm VHE}+\gamma_{\rm EBL}
\to e^+ e^-$)
on the extragalactic background
light (EBL) above roughly 100 GeV. 
In this respect, recent
observations of cosmologically distant gamma-ray sources by ground-based gamma-ray
Imaging Atmospheric Cherenkov Telescopes (IACTs) have revealed a surprising degree of transparency of the universe to 
VHE photons~\cite{Aharonian:2005gh,Meyer:2012}.
This result has been confirmed 
 by a  recent systematic analysis performed on 25 High-Energy photon sources, which 
 pointed out a suppression of the pair production during the propagation of VHE photons,
named ``pair production anomaly''~\cite{Horns:2012fx}.

Oscillations between VHE photons and ALPs can represent an intriguing -- but not unique, see e.g. \cite{steckershoks,Aharonian:2008su,Essey:2009zg,Essey:2009ju,Essey:2010er,Aharonian:2012fu}-- possibility to explain this anomaly.
Indeed, if VHE photons are transformed into a mixed photon-ALP state, the ALP component does not suffer from absorption effects while it propagates and can therefore reach the Earth from distant sources even at very-high energies. In
this sense, two complementary mechanisms have been proposed: a) VHE photon-ALP
conversions in the magnetic fields around gamma-ray 
sources and then further
back-conversions in the magnetic field of the Milky-Way~\cite{Simet:2007sa,Belikov:2010ma} ; b) oscillations
of VHE photons into ALPs in the turbulent extragalactic magnetic fields~\cite{De Angelis:2007dy,DMPR:2009,Mirizzi:2009aj,
SanchezConde:2009wu,Dominguez:2011xy,DeAngelis:2011id}.
Both these mechanisms are intriguing, but are affected by possible  drawbacks. 
In particular, concerning the mechanism a)  it is not clear at all whether a conversion $\gamma \to a$ actually takes place in all AGN mainly because their magnetic field is quite complicated and poorly 
known (see, e.g.,~\cite{SanchezConde:2009wu}).
On the other hand, mechanism b) requires intergalactic magnetic fields $B \sim 0.1-1$~nG
close to the current upper bounds, for which there is no firm observational evidence, or, alternatively, a very large $a\gamma\gamma$ coupling, at a level already excluded by other independent observations.%
\footnote{Recently, theoretical arguments have been discussed that the intergalactic medium is efficiently heated
through generation of plasma-instabilities by powerful blazars~\cite{Pfrommer:2011bg}. If this heating mechanism is at work,
it would imply a model-dependent upper limit on the  intergalactic magnetic field strength of $B \sim 10^{-3}$~nG.}

The aim of the present paper is to investigate in detail a third possibility concerning AGN hosted in clusters of galaxies, where a first conversion $\gamma \to a$ occurs in the magnetic field of the cluster while a reconversion $a \to \gamma$ happens in the magnetic field of the Milky Way. This scenario has  two important advantages. First, we are dealing with magnetic fields that are {\it known} to a considerable extent. Second, since a sizable fraction of the original photons travel all the time as ALPs, the overall EBL absorption is strongly reduced. As a consequence, a significant hardening of the VHE spectra of distant AGN located in galaxy clusters is expected.

The structure of the paper is as follows. In Sec.~2 we consider the observational evidence of AGN producing VHE photons in galaxy clusters. In Sec.~3 we review the formalism describing the photon-ALP mixing. In particular, we discuss the mechanism of photon-ALP oscillations in random magnetic fields which is relevant for $\gamma \to a$ conversions in galaxy clusters. In Sec.~4 we 
address the effect of photon-ALP oscillations on VHE photon spectra, describing the $\gamma \to a$ conversions in the intra-cluster magnetic fields, the absorption of photons in the intergalactic medium and the $a \to \gamma$ conversions in the Milky Way magnetic field. We  show how our mechanism can lead to a characteristic hardening of VHE photon spectra for distant sources at $E > 10 \, {\rm TeV}$. As a specific example we discuss in Sec.~5 the  effect on the energy spectra of two AGNs located in galaxy clusters, namely Mkn 501 at $z=0.034$~\cite{Aharonian:1999} and 1ES 0414+009 at redshift $z=0.287$~\cite{HESS:2012}. Finally, in Sec.~6 we  discuss future perspectives and we present our conclusions.

\section{Active Galactic Nuclei in Galaxy Clusters}
\label{agn:bfield}

The bulk of extra-galactic sources of VHE $\gamma$-rays has been identified with X-ray emitting AGN (see e.g.~\cite{wagner:2008}), which can be roughly described as an accretion disk around a supermassive black hole and two jets emanating from the centre and perpendicular to the disk. When one of the jet is directed towards us the AGN is called a Blazar. 

Blazars~\cite{urry} are divided into two broad groups: BL Lacs and Flat Spectrum Radio Quasars (FSRQs). BL Lacs are defined by the weakness of their thermal features like broad emission lines in their optical spectra. Accordingly, the nuclear region of BL Lacs, where the jet forms and accelerates is believed to be rather poor of soft photons. On the other hand, FSRQs display luminous broad emission lines, indicating the existence of photo-ionized clouds rapidly rotating around the central black hole and forming the so-called broad line region (BLR) at about one parsec from the centre.

BL Lacs are believed to be Fanaroff-Riley I type galaxies~\cite{wardle:1984,antonucci:1985,padovani:1990}. High spatial resolution optical observations indicate that the host galaxies of X-ray selected BL Lacs are typically giant elliptical galaxies \cite{wurtz:1996, urry:1999, falomo:1999}. Imaging observations of the environment of 45 Blazars up to $z=0.65$ indicate that the bulk of investigated objects avoid rich clusters~\cite{wurtz:1997}. However, according to a recent study of Sloan Digitial Sky Survey data including spectroscopic redshift information for objects with $z<0.4$, BL Lacs are found both in low and high density large scale environment
\cite{lietzen:2011}.  The sub-sample of VHE-emitting Blazars appears to populate both rich clusters as well as unspecific environments \cite{pepa:2012}. At this regard,   in Table~\ref{tab:sources} we present a sample of Blazars at different redshifts which have been found to be
 located in galaxy clusters. 
Moreover,  for a few Blazars not directly associated with a galaxy cluster, it can well happen that the line-of-sight traverses an intervening galaxy cluster.

\begin{table}[tbp]
\centering
\caption{A sample of Blazars in clusters of galaxies~\cite{tavecchioprivate,tevcat}.}

 \begin{tabular}{c|c|c}
  \hline
  \hline
  {Blazar} & {Celestial coordinates} & redshift\\
  
  \hline
  Mkn 501 & 16h53m52.2s + 39d45m37s & $z= 0.034$\\
  PKS 0548--322  &  05h50m38.4s -- 32d16m12.9s & $z=0.069$\\
  PKS 2005--489  & 20h09m27.0s -- 48d49m52s & $z=0.071$\\
  PKS 2155--304 & 21h58m52.7s -- 30d13m18s &  $z=0.116$\\
 1ES 1101--232 & 11h03m36.5s -- 23d29m31s & $ z=0.186$ \\
  1ES 0414+009 & 04h16m52.96s + 01d05m24s  & $z=0.287$ \\
  \hline
 \end{tabular}
\label{tab:sources}
\end{table}

The existence of magnetic fields in galaxy clusters is well established through the observation of radio synchrotron emission as well as through the rotation measure of polarized radio sources (for a review, see e.g.~\cite{carilli:2002}). The structure of the magnetic field within galaxy clusters has been subject to a number of studies, most notably Faraday rotation measurements and subsequent modeling of the underlying magnetic field structure~\cite{carilli:2002,govoni:2004}. For well-observed objects like e.g. the Coma and Hydra A clusters, a Kolmogorov-type power spectrum has been found to fit the data for scales up to tens of kpc~\cite{bonafede:2010, kuchar:2011} with field strengths of a few up to ten $\mu$G in the inner range ($<100$~kpc) of the cluster. The cluster magnetic field connects very likely smoothly to magnetic fields on larger scales (filament), which may be close to $\mu$G strength according to simulations~\cite{dolag:2005,brueggen:2005}.

 This motivates our choice of parameters throughout this paper: A field strength $B=1~\mu$G and a coherence length $l_c=10$~kpc. According to the standard lore, the intra-cluster magnetic field ${\bf B}$ is modeled as a network of magnetic domains with a size equal to the coherence length. In every domain ${\bf B}$ is assumed to have the same strength but its direction is allowed to change randomly from one domain to another.
The typical electron density in the intra-cluster medium is \mbox{$n_e \simeq 1.0 \times 10^{-3} \, {\rm cm}^{-3}$}~\cite{sarazin1986}.

\section{Oscillations of photons into axion-like particles}

\subsection{General considerations}

Photon-ALP mixing occurs in the presence of an external magnetic field ${\bf B}$ due to the interaction term~\cite{Raffelt:1987im,sikivie,Anselm:1987vj}
\begin{equation}
\label{mr}
{\cal L}_{a\gamma}=-\frac{1}{4} \,\gag
F_{\mu\nu}\tilde{F}^{\mu\nu}a=\gag \, {\bf E}\cdot{\bf B}\,a~,
\end{equation}
where $\gag$ is the photon-ALP coupling constant (which has the dimension of an inverse energy). 

We consider throughout a monochromatic photon/ALP beam of energy $E$ propagating along the $x_3$ direction in a cold ionized and magnetized medium. 
It has been shown that for very relativistic ALPs and {\it polarized} photons,
 the beam
propagation equation can be written in a Schr\"odinger-like form in which $x_3$
takes the role of time~\cite{Raffelt:1987im}
\begin{equation}
\label{we} 
\left(i \, \frac{d}{d x_3} + E +  {\cal M} \right)  \left(\begin{array}{c}A_{1} (x_3) \\ A_2 (x_3) \\ a (x_3) \end{array}\right) = 0~,
\end{equation}
where $A_1(x_3)$ and $A_2 (x_3)$ are the  photon linear polarization amplitudes along the $x_1$ and $x_2$ axis, respectively, $a (x_3)$ denotes the ALP amplitude  and ${\cal M}$ represents the photon-ALP mixing matrix. 
We denote by $T (x_3,0;E)$ the transfer function, namely the solution of Eq. (\ref{we}) with initial condition $T (0,0;E) = 1$. 

The mixing matrix ${\cal M}$  simplifies if we restrict our attention to the case in which ${\bf B}$ is homogeneous. We denote by ${\bf B}_T$ the transverse magnetic field, namely its component in the plane normal to the beam direction and we choose the $x_2$-axis along ${\bf B}_T$ so that $B_1$ vanishes. The linear photon polarization state parallel to the transverse field direction ${\bf B}_T$ is then denoted by $A_{\parallel}$ and the orthogonal one by $A_{\perp}$. Correspondingly, the mixing matrix can be written as~\cite{Mirizzi:2005ng,Mirizzi:2006zy}
\begin{equation}
{\cal M}_0 =   \left(\begin{array}{ccc}
\Delta_{ \perp}  & 0 & 0 \\
0 &  \Delta_{ \parallel}  & \Delta_{a \gamma}  \\
0 & \Delta_{a \gamma} & \Delta_a 
\end{array}\right)~,
\label{eq:massgen}
\end{equation}
whose elements are~\cite{Raffelt:1987im} $\Delta_\perp \equiv \Delta_{\rm pl} + \Delta_{\perp}^{\rm CM},$ $ \Delta_\parallel \equiv \Delta_{\rm pl} + \Delta_{\parallel}^{\rm CM},$ $\Delta_{a\gamma} \equiv {g_{a\gamma} B_T}/{2} $ and $\Delta_a \equiv - {m_a^2}/{2E}$, where $m_a$ is the ALP mass. The term $\Delta_{\rm pl} \equiv -{\omega^2_{\rm pl}}/{2E}$ accounts for plasma effects, where $\omega_{\rm pl}$ is the plasma frequency expressed as a function of the electron density in the medium $n_e$ as $\omega_{\rm pl} \simeq 3.69 \times 10^{- 11} \sqrt{n_e /{\rm cm}^{- 3}} \, {\rm eV}$. The  terms $\Delta_{\parallel,\perp}^{\rm CM}$ describe the Cotton-Mouton  effect, i.e.~the birefringence of fluids in the presence of a transverse magnetic field.  A vacuum Cotton-Mouton effect is expected from QED one-loop corrections to the photon polarization in the presence of an external magnetic field $\Delta_\mathrm{QED} = |\Delta_{\perp}^{\rm CM}- \Delta_{\parallel}^{\rm CM}| \propto B^2_T$, but this effect is completely negligible in the present context. An off-diagonal $\Delta_{R}$ would induce the Faraday rotation, which is however totally irrelevant at VHE, and so it has been dropped. For the relevant parameters,  we numerically find 
\begin{eqnarray}  
\Delta_{a\gamma}&\simeq &   7.6\times10^{-2} \left(\frac{g_{a\gamma}}{5\times 10^{-11}\textrm{GeV}^{-1}} \right)
\left(\frac{B_T}{10^{-6}\,\rm G}\right) {\rm kpc}^{-1}
\nonumber\,,\\  
\Delta_a &\simeq &
 -7.8 \times 10^{-3} \left(\frac{m_a}{10^{-8} 
        {\rm eV}}\right)^2 \left(\frac{E}{{\rm TeV}} \right)^{-1} {\rm kpc}^{-1}
\nonumber\,,\\  
\Delta_{\rm pl}&\simeq & 
  -1.1\times10^{-10}\left(\frac{E}{{\rm TeV}}\right)^{-1}
         \left(\frac{n_e}{10^{-3} \,{\rm cm}^{-3}}\right) {\rm kpc}^{-1}
\nonumber\,,\\
\Delta_{\rm QED}&\simeq & 
4.1\times10^{-6}\left(\frac{E}{{\rm TeV}}\right)
\left(\frac{B_T}{10^{-6}\,\rm G}\right)^2 {\rm kpc}^{-1}
\,. 
\label{eq:Delta0}\end{eqnarray}

For the above estimates that we will use in the following as benchmark values, we refer to the following physical inputs: The strength of $B$-fields and the electron density $n_e$ are typical values for galaxy clusters mentioned in Section~\ref{agn:bfield}, the value of the photon-ALP coupling $g_{a \gamma}$ is below the direct experimental bound $g_{\rm a\gamma}\lesssim 8.8\times 10^{-11}$~GeV$^{-1}$ obtained by the CAST experiment for 
$m_a\lesssim 0.02$~eV~\cite{Arik:2011rx}, slightly better than the long-standing globular-cluster limit~\cite{Raffelt:2006cw}. We recall that for ultra-light ALPs ($m_a \lesssim 10^{- 10}$~eV) a more stringent limit $\gag\lesssim 1 \times 10^{-11}$~GeV$^{-1}$~\cite{Brockway:1996yr} or even $\gag \lesssim 3\times 10^{-12}$~GeV$^{-1}$~\cite{Grifols:1996id} has been derived from the absence of $\gamma$-rays from SN~1987A even if with a large uncertainty.  
Another strong bound, namely $\gag \lesssim  10^{-11}$~GeV$^{-1}$ for $m_a \lesssim 10^{- 7}$~eV has been recently presented
in~\cite{Gill:2011yp}, based on the photon polarization in magnetic white dwarfs. 
However, since it is based on a simplified model of the white dwarf environment in which photons propagate, 
it should be taken only as indicative. 
In general we believe that since
every experimental measurement and every astrophysical 
argument has its own systematic uncertainties and its own recognized or un-recognized
loop holes,  to corner ALPs it is certainly important to use as many
independent interaction channels and as many different approaches as possible.

\subsection{Single magnetic domain}

Considering the propagation of photons in a single magnetic domain with a uniform ${\bf B}$-field with $B_1 = 0$, the  component $A_{\perp}$ decouples away, and the propagation equations reduce to a 2-dimensional problem. Its solution follows from the diagonalization of the 2-dimensional mixing sub-matrix of ${\cal M}_0$ through a similarity transformation performed with an orthogonal matrix, parametrized by the rotation angle $\theta$ which takes the value~\cite{Raffelt:1987im}
\begin{equation}
\theta = \frac{1}{2}\textrm{arctan}\left(\frac{2 \Delta_{a \gamma}}{\Delta_{\rm pl}-\Delta_{a}}\right) \,\ .
\label{theta}
\end{equation}
In particular, the probability for a photon emitted in the state $A_{\parallel}$ to oscillate into an ALP after traveling a distance $d$ is given by~\cite{Raffelt:1987im}
\begin{eqnarray}
P_{\gamma \to a}^{(0)}  &=& {\rm sin}^2 2 \theta \  {\rm sin}^2
\left( \frac{\Delta_{\rm osc} \, d}{2} \right)  \,\ \nonumber \\
&=& (\Delta_{a \gamma} d)^2 \frac{\sin^2(\Delta_{\rm osc} d/2)}{(\Delta_{\rm osc} d/2)^2} \,\ ,
\label{conv}
\end{eqnarray}
where the oscillation wave number is
\begin{equation}
\Delta_{\rm osc} \equiv \left[(\Delta_{a} - \Delta_{\rm pl})^2 + 4 \Delta_{a \gamma}^2 \right]^{1/2}~.
\end{equation}
It proves useful to define a critical energy~\cite{De Angelis:2007yu}
\begin{eqnarray}
{E}_c &\equiv& 
\frac{E |\Delta_a-\Delta_{\rm pl}|}{2 \Delta_{a \gamma}}  \nonumber\\
&\simeq&   
\frac{50 | m_a^2 - {\omega}_{\rm pl}^2|}{(10^{-8}{\rm eV})^2}
\left( \frac{10^{-6}{\rm G}}{B_T} \right)
\left( \frac{5\times10^{-11}\rm GeV^{-1}}{g_{a \gamma}} \right)
{\rm GeV}~. \nonumber
\label{eq:Ecrit}
\end{eqnarray}
in terms of which the oscillation wave number can be rewritten as
\begin{equation}
\label{a17t}
{\Delta}_{\rm osc} = 2 \Delta_{a \gamma} \sqrt{1+ \left(\frac{E_{\rm c} }{E} \right)^2}~.
\end{equation}
From Eqs.~(\ref{theta}) -- (\ref{a17t}) it follows that in the energy range $ E \gg E_{\rm c}$ the  photon-ALP mixing is maximal ($\theta \simeq \pi/4$) and the conversion probability becomes energy-independent. This is the so-called {\it strong-mixing regime}. Outside this regime the conversion probability turns out to be energy-dependent and vanishingly small, so that $E_{\rm c}$ acquires the meaning of a  low-energy  cut-off. 

So far, we have been dealing with a beam containing polarized photons, but since at VHE the polarization cannot be measured we better assume that the beam is unpolarized. This is properly done by means of the polarization density matrix 
\begin{equation}
\rho (x_3) = \left(\begin{array}{c}A_1 (x_3) \\ A_2 (x_3) \\ a (x_3)
\end{array}\right)
\otimes \left(\begin{array}{c}A_1 (x_3)\  A_2 (x_3)\ a (x_3)\end{array}\right)^{*}
\end{equation}
which obeys the Liouville-Von Neumann equation~\cite{Bassan:2010ya}
\begin{equation}
\label{vne}
i \frac{d \rho}{d x_3} = [\rho, {\mathcal M_0}]
\end{equation}
associated with Eq. (\ref{we}). Then it follows that the solution of Eq. (\ref{vne}) is given by 
\begin{equation}
\label{t5L}
\rho (x_3,E) = T (x_3, 0;E) \, \rho (0) \, T^{\dagger}(x_3, 0;E)~, 
\end{equation}
where $\rho (0)$ is the initial beam state. Note that for a uniform ${\bf B}$ even if we clearly have
\begin{equation}
\label{t5Lq}
T (x_3, 0;E) = e^{i(E + \mathcal{M}_0) x_3}~, 
\end{equation}
Eq. (\ref{t5L}) reads
\begin{equation}
\label{t5Lw}
\rho (x_3,E) = e^{i \mathcal{M}_0 \, x_3} \, \rho (0) \, e^{- i \mathcal{M}_0 \, x_3}~. 
\end{equation}

\subsection{Network of domains with random magnetic fields}

Since in the following we will consider VHE photons emitted by an AGN in a galaxy cluster, we have to deal with a more general situation than the one depicted in the previous Section. Indeed, 
as discussed in Sec.~\ref{agn:bfield}, the intra-cluster ${\bf B}$-field has a domain-like structure with size set by its coherence length. Although the strength of ${\bf B}$ is supposed to be the same in every domain its direction changes randomly from one domain to another. Therefore the propagation over many magnetic domains is clearly a truly 3-dimensional problem, because -- due to the randomness of the direction of ${\bf B}$ -- the mixing matrix ${\cal M}$ entering the beam propagation equation cannot be reduced to a block-diagonal form similar to ${\cal M}_0$ in all domains. Rather, we take the $x_1$, $x_2$, $x_3$ coordinate system as fixed once and for all, and -- denoting by $\psi_k$ the angle between $B_T$ and the $x_2$ axis in the generic $k$-th domain ($1 \leq k \leq n$) -- we treat every $\psi_k$  as a random variable in the range $0 \leq \psi_k < 2 \pi$. During their path with a total length $L$ in the galaxy cluster, the beam crosses $n = L/l_c$ domains, where $l_c$ is the size of each domain: The set $ \{B_k \}_{1 \leq k \leq n}$ represents a given random realization of the beam propagation corresponding to the set $ \{\psi_k \}_{1 \leq k \leq n}$. Accordingly, in each domain the matrix ${\cal M}$ takes the form~\cite{Mirizzi:2005ng}
 \begin{equation}
\label{aa8MR}
{\cal M}_k = \left(
\begin{array}{ccc}
\Delta_{xx} & \Delta_{xy} & \Delta_{a\gamma} \, \sin\psi_k\\
\Delta_{yx} & \Delta_{yy} & \Delta_{a\gamma} \, \cos\psi_k\\
\Delta_{a\gamma} \, \sin\psi_k& \Delta_{a\gamma} \, \cos\psi_k& \Delta_{a} \\
\end{array}
\right)~,
\end{equation} 
with 
\begin{equation}
\Delta_{xx} = \Delta_\parallel \, \sin^2 \psi_k + \Delta_\perp \cos^2 \psi_k~,
\end{equation}
\begin{equation}
\Delta_{xy} = \Delta_{yx}=(\Delta_\parallel -\Delta_\perp) \sin\psi_k \, \cos\psi_k~,
\end{equation}
\begin{equation}
\Delta_{yy} = \Delta_\parallel \cos^2 \psi_k + \Delta_\perp \sin^2 \psi_k~.
\end{equation}

Working in terms of the Eq. (\ref{vne}), after the propagation over $n$ magnetic domains the density matrix is given by repeated use of Eq. (\ref{t5Lw}) with $M_0 \to M_k$, namely~\cite{Bassan:2010ya}
\begin{equation}
\label{mr1}
\rho_n = T (\psi_n, \ldots , \psi_1) \, \rho_0 \, T^{\dagger} (\psi_n, \ldots , \psi_1)~,
\end{equation}
where we have set
\begin{equation}
T (\psi_n, \ldots , \psi_1)  \equiv \prod^n_{k = 1} T_k~,
\end{equation}
 with
\begin{equation}
T_k = e^{i {\cal M}_k l_c} \,\ ,
\end{equation}
which is the transfer function in the $k$-th domain (as explained above, the $E$-dependent factor drops out from Eq. (\ref{mr1}) and so it has been neglected). 
 
Since we  do not  know the particular configuration crossed by the beam during its propagation, in order to get an idea of the effect induced by the $\gamma \to a$ conversions it is useful to perform an ensemble average over all the possible realizations encompassing the $1, \ldots n$ domains. Assuming that the conversion probability $P_{a \to \gamma}^{(0)}$ in each magnetic domain 
[Eq.~(\ref{conv})] is small, the average photon flux after $n$ domains reads
\begin{equation}
I_{\gamma}^n = {\bar\rho}_{11} + {\bar\rho}_{22} = P_{\gamma \to \gamma}^{\rm CL} I_{\gamma}^0~,
\label{eq:flux}
\end{equation}
where $ I_{\gamma}^0$ is the emitted photon flux. For an initial unpolarized photon state 
\begin{equation}
\rho_0 = \frac 1 2 \mathrm{diag}(1,1,0)
\label{eq:massgenA}
\end{equation}
we obtain~\cite{Mirizzi:2006zy}
\begin{equation}
P_{\gamma \to \gamma}^{\rm CL}=  \frac{2}{3} + \frac{1}{3}\left(1-\frac{3}{2}P_{a \to \gamma}^{(0)}\right)^n~,
\label{eq:probclust}
\end{equation}
where the ALP-photon conversion probability $P_{a \to \gamma}^{(0)}$ is provided by Eq.~(\ref{conv}).

\section{The mechanism}

Equipped with the results of the previous Section, we are now ready to investigate the implications of the mechanism we are considering for the conversions $\gamma \to a \to \gamma$ in the VHE  range.
More specifically, we first address the $\gamma \to a$ conversions in a galaxy cluster (Sec.~\ref{subsec:intra}), then the 
propagation of photons in the intergalactic medium where pair-production effects are relevant (Sec.~\ref{subsec:extra}) and finally the back-conversions $a \to \gamma$ inside the Milky Way (Sec.~\ref{subsec:MW}).

For definiteness, we will assume in the following a photon-ALP coupling $g_{a\gamma}= 5 \times 10^{-11}$~GeV$^{-1}$ and an ALP mass $m_a = 10^{-8}$~eV. With these values, the critical energy in Eq.~(\ref{eq:Ecrit}) above which the strong-mixing regime in a galaxy cluster takes place is $E_c \simeq 50$~GeV. We remark that for our input parameters, 
 we can neglect $\gamma \leftrightarrow a$ conversions in the intergalactic medium. Indeed, 
assuming a strength of the intergalactic magnetic field $B \sim 1$~nG as required for sizable photon-ALP conversions, the critical energy is $E_c \simeq 500$~TeV, far above the reach of the experimental detectability. Conversions in the intergalactic
$B$-fields 
 would otherwise be relevant
for values of $m_a \lesssim 10^{- 10} \, {\rm eV}$. Finally, since our goal is to understand the relevance of galaxy clusters for $\gamma \to a$ conversion, we assume that no such conversion occurs inside the Blazar.

\subsection{ Conversions in intra-cluster magnetic fields} 
\label{subsec:intra}

\begin{figure}[!t]  
\includegraphics[angle=0,width=1.\columnwidth]{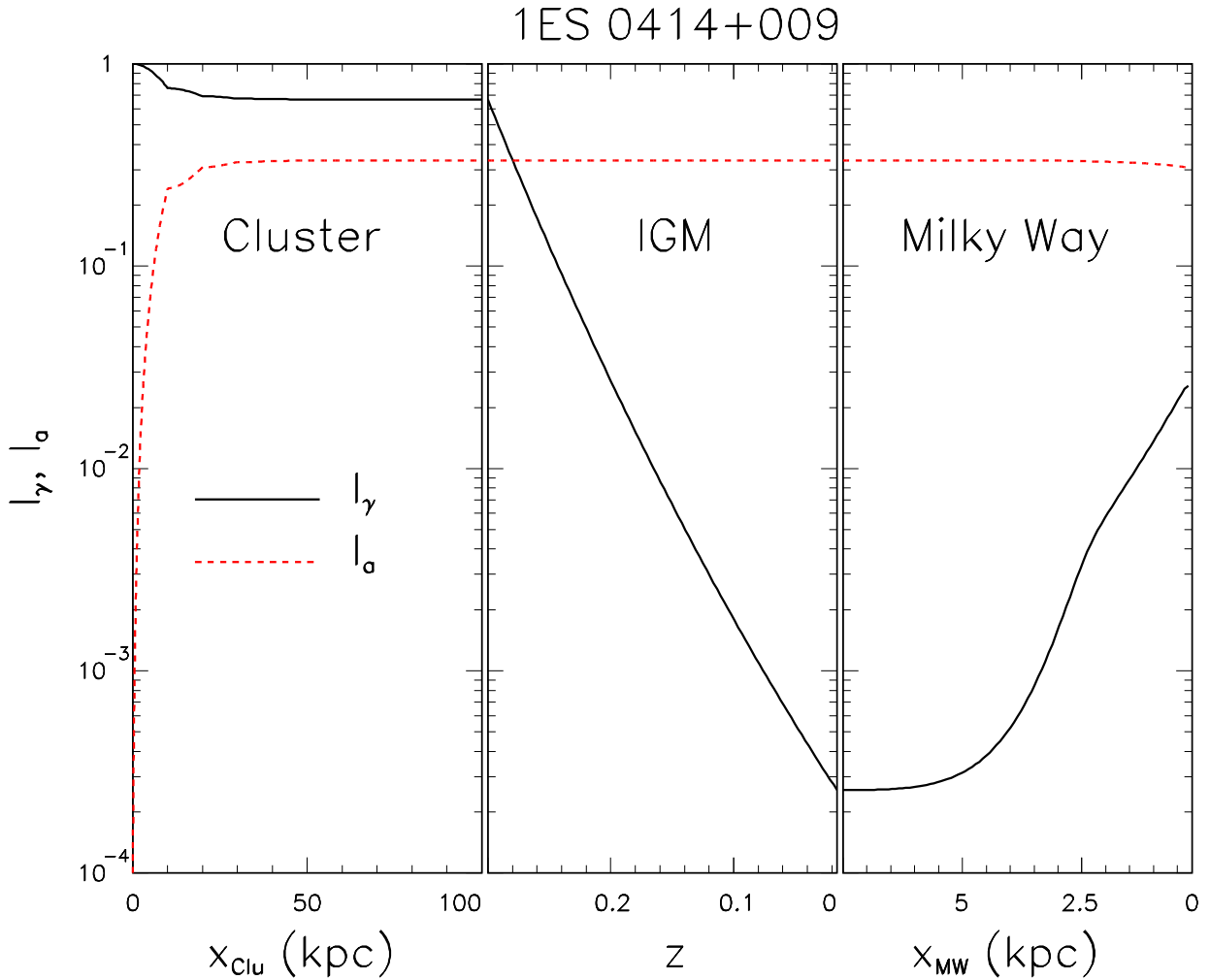} 
 \caption{Evolution of the photon flux $I_{\gamma}$ and ALP flux $I_{a}$  (the emitted photon flux is normalized to 1) for the Blazar 1ES 0414+009 at redshift  $z=0.287$ for a representative energy $E=8$~TeV. The left panel represents the evolution inside the galaxy cluster, the central one the evolution in the intergalactic medium (photon absorption and ALP free propagation, with redshift on the horizontal axis), and the right one the evolution within the Milky Way.}
\label{fig1}
\end{figure}

As discussed in Sec.~\ref{agn:bfield}, we assume a cellular structure for the intra-cluster magnetic fields, with domain size \mbox{$l_c \simeq 10$~kpc}. Moreover, for \mbox{$n_e \simeq 1.0 \times 10^{-3} \, {\rm cm}^{-3}$} the plasma frequency turns out to be \mbox{${\omega}_{\rm pl} \simeq 1.2 \times 10^{-12} \, {\rm eV}$}. With these input values we are in the strong-mixing regime, where the photon-ALP conversion probability is energy-independent. From Eqs.~(\ref{eq:flux}) -- (\ref{eq:probclust}) the average flux of photons and ALPs coming out from the galaxy cluster are given by

\begin{eqnarray}
I_{\gamma}^{\rm CL} & = & P_{\gamma \to \gamma}^{\rm CL} I_{\gamma}^0~, \nonumber \\
I_{a}^{\rm CL} & = & \Bigl(1-P_{\gamma \to \gamma}^{\rm CL} \Bigr) I_{\gamma}^0~.
\label{fluxclust}
\end{eqnarray}
The evolution of these fluxes inside the galaxy cluster -- with $I_{\gamma}^0$ normalized to 1 -- are represented in the left panel of 
Fig.~\ref{fig1}, considering the case of Blazar 1ES 0414+009 for a representative energy $E=8$~TeV.
It turns out that on average the fraction of VHE photons converting into ALPs within the cluster is $\sim 30~\%$. However, significant variations are possible along a given line of sight, owing to the random nature of the intra-cluster magnetic field. In particular, it is straightforward to show that for an initially unpolarized photon beam the ALP flux coming out of the cluster can vary between $0 \%$ and 
$50 \%$.

\subsection{Absorption of VHE photons on extragalactic background light}
\label{subsec:extra}

The ALPs produced in a galaxy cluster propagate over cosmological distance undisturbed until they reach the Galaxy. On the other hand, VHE photons undergo absorption due to the pair-production process off EBL low energy photons $\gamma^{\rm VHE}\gamma^{\rm EBL}\to e^+e^-$. The energy range 100~GeV $\lesssim E \lesssim$ 10~TeV relevant for presently operating IACTs absorption is dominated by the interactions with optical/infrared EBL photons. The absorption rate  for such a process as a function of  the incident VHE photon energy $E$ is~\cite{pp3,pp4,pp5}
\begin{equation}
\Gamma_\gamma(E)=\int_{m_e^2/E}^{\infty} d\epsilon ~ \frac{d n^{\rm EBL}_\gamma}{d\epsilon}
\int_{-1}^{1-\frac{2m_e^2}{E\epsilon}} d\xi ~ \frac{1-\xi}{2} \, \sigma_{\gamma\gamma}(\beta) \,\ ,
\, \label{eq:gamma}\end{equation}
where $\epsilon$ and $n^{\rm EBL}_\gamma$ are the EBL photon energy and number density, respectively, $\xi$ is the cosine of the angle between the VHE and the EBL photon momenta and the limits of integration in both integrals are determined by the kinematic threshold of the process. Moreover
\begin{equation}
\nonumber
\sigma_{\gamma\gamma}(\beta) = 1.25 \times 10^{-25} \left(1-\beta^2 \right) \left[2\beta \left(\beta^2-2 \right)+ \left(3-\beta^4 \right) \log\frac{1+\beta}{1-\beta}\right] \, {\rm cm}^2~,
\end{equation}
is the cross-section for the pair-production process~\cite{pp1,pp2} as a function of the electron velocity in the center-of-mass frame 
$\beta=[1-2m_e^2/E\epsilon(1-\xi)]^{1/2}$.

Several realistic models for the EBL are available in the literature which rely upon different strategies
(see, e.g.,~\cite{primack3,gilmore, Franceschini:2008tp,Franceschiniweb,finke, dominguez,mazinraue5,dwek2011,
Kneiske:2010,Kneiske:2010web}). Remarkably, they are basically in agreement with each other. 
Among the possible choices, we will employ as benchmark EBL  the Franceschini-Rodighiero-Vaccari (FRV) model~\cite{Franceschiniweb,Franceschini:2008tp}. We will also compare our results with the minimal EBL Kneiske model~\cite{Kneiske:2010,Kneiske:2010web}, which provides a strict lower-limit for the attenuation of VHE $\gamma$-rays. As we will see, our conclusions will be rather independent of the choice of the EBL model, as the dominant part of the $\gamma$-ray flux at Earth is mostly given by back-conversions $a \to \gamma$, as it is already clear from our Fig.~\ref{fig1}.

According to conventional physics, for a given Blazar at distance $d_0$ (corresponding to a redshift $z_0$), the photon spectrum observed at Earth is given by
\begin{equation}
I^{E}_\gamma(E) = \exp\left(-\tau_\gamma\right) 
I^0_\gamma(E_0) \, ,
\label{eq:noconversion}\end{equation} 
where $I^0(E_0)$ is the emitted spectrum with initial photon energy $E_0=E(1+z_0)$ and $\tau_\gamma$ is the {\em optical depth} which accounts for the EBL absorption and reads  
\begin{eqnarray}
\tau_\gamma &=&\int_0^{d_0} dx \,\Gamma_\gamma(E,x) \nonumber \\
&=& \frac{c}{H_0}\int_{0}^{z_0}\frac{dz \,\ \Gamma_{\gamma}(E,z) }{(1+z) 
\sqrt{\Omega_\Lambda + \Omega_{\rm m}(1+z)^3}} \,\ ,
\end{eqnarray}
where $H_0=73$~km~Mpc$^{-1}$~s$^{-1}$ is the Hubble constant, $\Omega_{\rm m}=0.24$~\cite{pdg} is the matter density and 
$\Omega_\Lambda =1-\Omega_{\rm m}$ is the dark energy density (assuming a flat cosmology).

Instead in the presence of $\gamma \to a$ conversions in galaxy clusters, only the photons coming out  of the cluster will be EBL absorbed, as shown in the central panel of Fig.~\ref{fig1}. Therefore, the average fluxes reaching the edge of the Milky Way turn out to be
\begin{eqnarray}
I_{\gamma}^{\rm MW} & = &\exp\left(-\tau_\gamma\right) I_{\gamma}^{\rm CL} =
\exp\left(-\tau_\gamma\right) P_{\gamma \to \gamma}^{\rm CL} I_{\gamma}^0 \,\ , \nonumber \\
I_{a}^{\rm MW} & = & I_{a}^{\rm CL} = \Bigl(1-P_{\gamma \to \gamma}^{\rm CL} \Bigr) I_{\gamma}^0 \,\ .
\end{eqnarray}
Given the strong attenuation of the photon flux due to the EBL absorption, we get $I_{\gamma}^{\rm MW} \ll I_{a}^{\rm MW}$.

\subsection{Back-conversions in the Milky Way}
\label{subsec:MW}

Ultimately the photon/ALP beam crosses the Milky Way before being detected. Observations over the last three decades have led to a rather detailed picture of the Milky Way magnetic field. Perhaps, its most important feature is that it consists in two components, a regular and a turbulent one. The latter component can be described by a cellular structure, with strength $B \simeq 1 \times 10^{-6}$~G
and domain size $l_{\rm MW} \simeq 10^{-2}$~pc~\cite{ferriere}. It is straightforward to realize that in this case the
oscillation length $l_{\rm osc} = 2\pi/\Delta_{\rm osc}$ is much larger than the domain size $l_{\rm MW}$, so that the
$a \to \gamma$ conversion is vanishingly small. Therefore, in the following we restrict our attention to the regular component which is relevant for $a \to \gamma$ conversions.

Measurements of the Faraday rotation based on pulsar observations have shown that this component is parallel to the Galactic plane, apart from a possible dipole component at the galactic center and a small vertical component. Its strength varies between $B \simeq 1.4 \times 10^{- 6} \, {\rm G}$ in the Solar neighborhood and $B \simeq 4.4 \times 10^{- 6} \, {\rm G}$ in the inner Norma arm \cite{han,Beck:2008ty}. Moreover, the associated radial coherence length is $l_r \simeq10 \, {\rm kpc}$ \cite{Beck:2008ty}, while the vertical scale height is of the order of a few kpc. Inside the Milky Way disk the electron density is \mbox{$n_e \simeq 1.1 \times 10^{-2} \, {\rm cm}^{-3}$}~\cite{Digel}, resulting in a plasma frequency ${\omega}_{\rm pl} \simeq 4.1 \times 10^{-12} \, {\rm eV}$.

The regular component of the galactic magnetic field is typically split into a disk and a halo part. The disk component is most often modeled as a logarithmic spiral with either symmetric or antisymmetric behavior with respect to the galactic plane (see e.g.~\cite{Sun:2007mx,Jansson:2009ip} and references therein). Additionally, depending on whether the direction of the field in two different arms is the same or opposite, the model is called axisymmetric (ASS, also dubbed disymmetric, DSS, in \cite{Jansson:2009ip}) or bisymmetric (BSS), respectively. Its typical scale height is of order 1 kpc. The halo component is instead typically taken as a purely toroidal (i.e.~azimuthal) field \cite{Prouza:2003yf}, with a slightly larger vertical scale height, possibly extending up to $\gtrsim$ 3 kpc. The halo field can be different above and below the galactic plane. 

Several models have been proposed in the literature, based on combined analyses of Faraday rotation measurements of extra-galactic sources and of polarized galactic diffuse radio emission. However, statistical fits of these models to data are generally poor \cite{Men:2008fm,Jansson:2009ip}. Recently, evidence for an extra, out-of-plane, ``$X$-shaped'' component has been suggested \cite{Jansson:2012pc}, together with an indication of large vertical scale height of the order of 5 kpc for the halo field component. While we take the Jansson and Farrar model \cite{Jansson:2012pc} as our benchmark, we also consider the best fit model derived by Pshirkov et al. \cite{Pshirkov:2011um} (see their table 3) from a global analysis of Faraday rotation measurements of extragalactic sources, to show the systematics associated to the poor knowledge of galactic magnetic fields. We further check some of the models discussed in~\cite{Jansson:2009ip}, which give results similar to the ones of~\cite{Pshirkov:2011um}. 

\begin{figure}[tbp]
\centering
\includegraphics[scale=0.4]{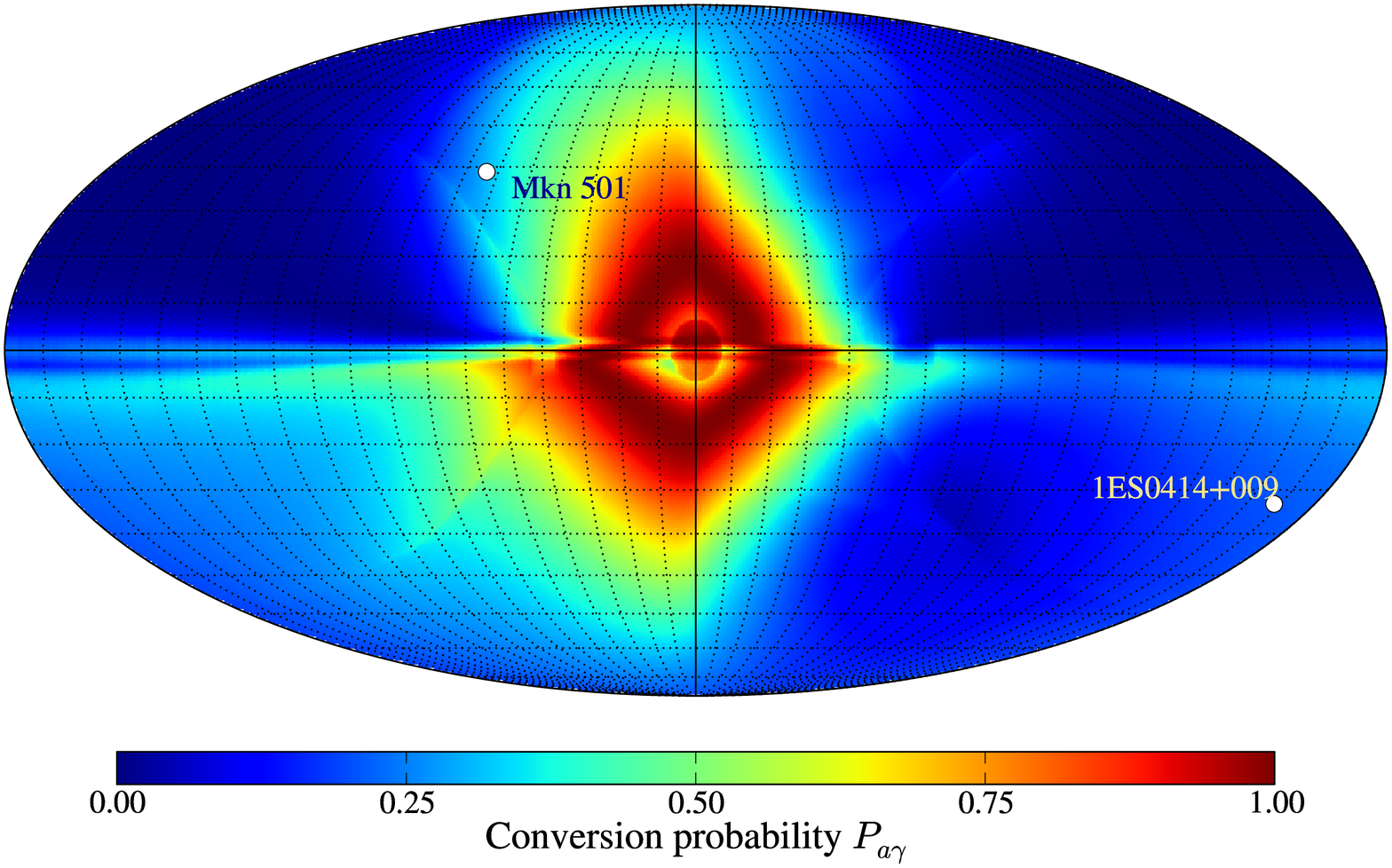}
\includegraphics[scale=0.4]{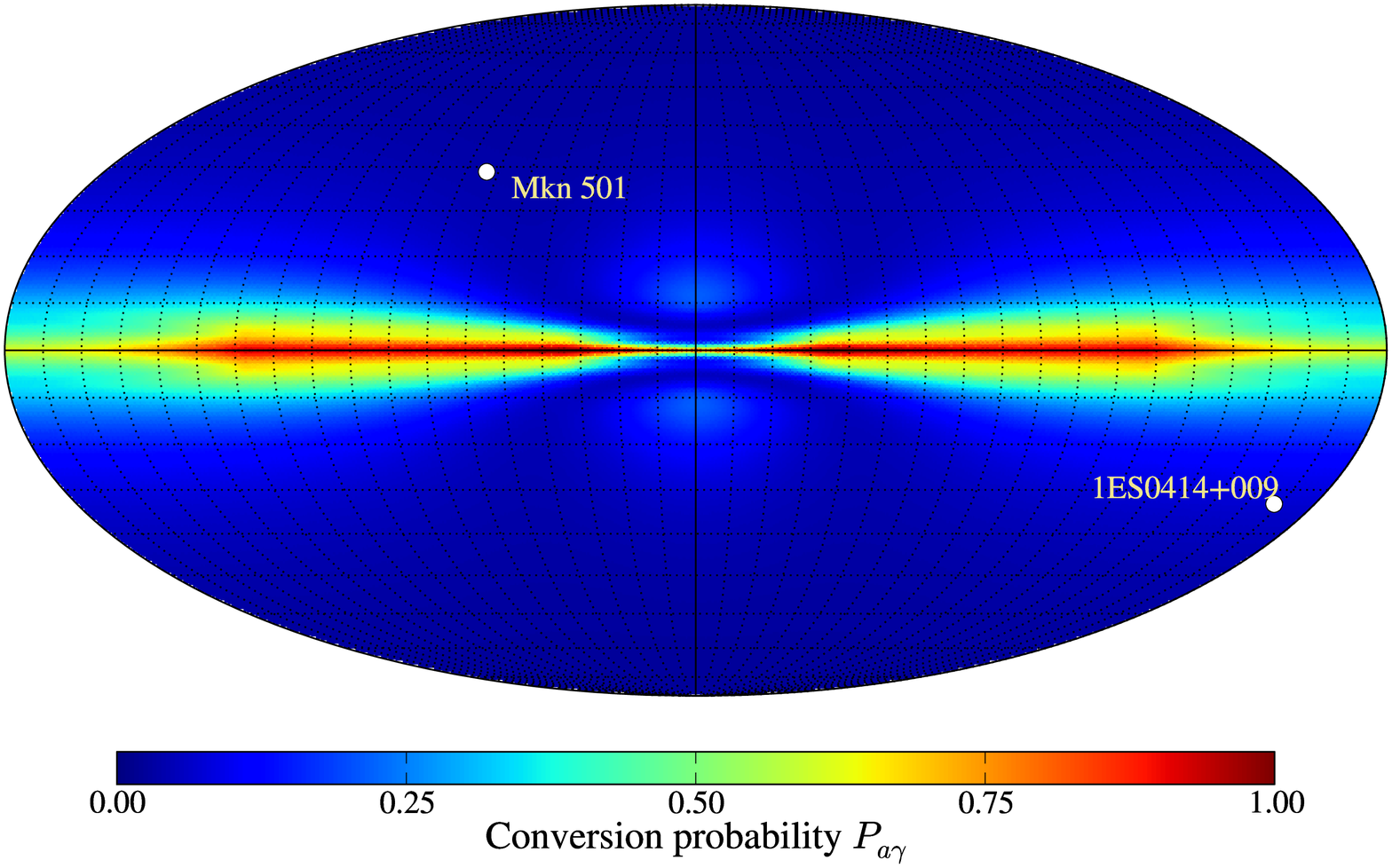}
\caption{Skymap in galactic coordinates of the $a \to \gamma$ conversion probability, starting from a pure ALPs beam at the outside boundary of the Galaxy, for the  Jansson and Farrar magnetic field model derived in \cite{Jansson:2012pc} (upper panel) and the one of Pshirkov et al.  \cite{Pshirkov:2011um} (lower panel), energy $E = 8~{\rm TeV}$, coupling $g_{a\gamma} = 5\times10^{-11}~{\rm GeV}^{-1}$ and $m_{a}=10^{-8}~{\rm eV}$. We also show the sky position of the two Blazars 1ES 0414+009 and Mkn 501.}
\label{fig:conversionmap}
\end{figure}

In order to compute the $a \to \gamma$ conversions in the Galaxy, we integrate Eq.~(\ref{vne}) along each given galactic line of sight inside it. An illustrative sky map of the line-of-sight dependent probability for an ALP at the edge of the Galaxy to convert into a photon at Earth is shown in Fig.~\ref{fig:conversionmap}  for our chosen reference
Jansson and Farrar magnetic field model (upper panel) and for the Pshirkov et al. model~\cite{Pshirkov:2011um} (lower panel). The effect of the different magnetic field models is striking. In the case of Jansson and Farrar
model the probability of $a \to \gamma$ conversion is generally larger than for the Pshirkov et al., especially towards the galactic center and in the southern galactic halo, due to the presence of the $X$-shaped field and to the large vertical scale height of the halo field.  This would favor the detection of hardened VHE spectra from extragalactic sources if the mechanism we are proposing is indeed at work.

The position of the two Blazars under consideration is also marked in Fig.~\ref{fig:conversionmap}.
Both sources are located  in a region where the back-conversion probability can be of the order of a few percent up to $\sim30\%$. We remark that the two sources considered here are located in directions where different magnetic field models give comparable results, within a factor of a few, for the conversion probability.

\begin{figure}[!t]  
\includegraphics[angle=0,width=1.\columnwidth]{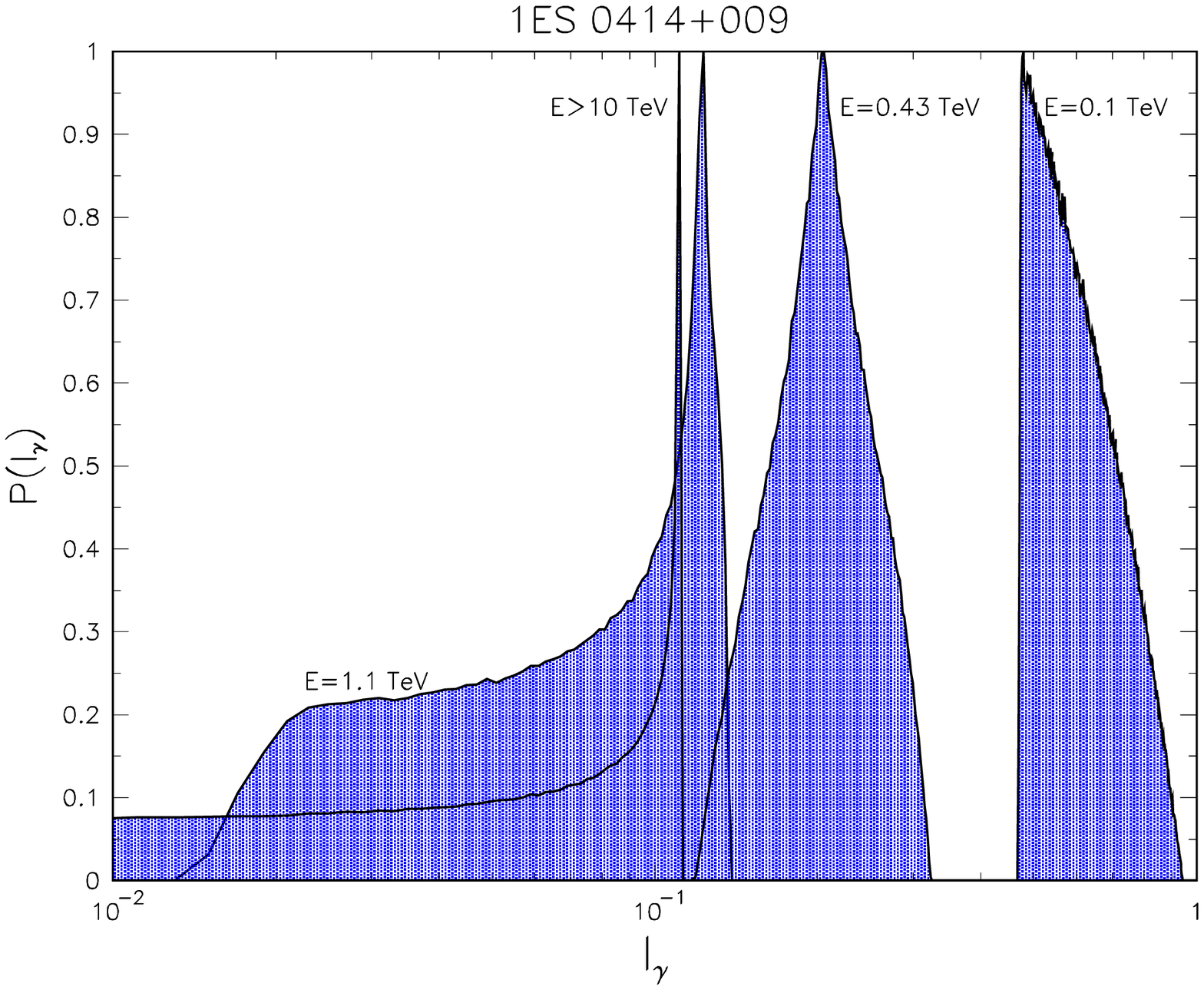} 
 \caption{Probability distribution functions (normalized to the maximum) for the observable photon flux $I_{\gamma}$ from 
 1ES0414+009 at different energies.}\label{fig3} 
\end{figure}  

Moreover, as we already explained, we show in the right panel of Fig.~\ref{fig1} the evolution of the photon and ALP fluxes along the line-of-sight to the source 1ES 0414+009 in the Milky-Way. The observable average VHE photon flux at Earth is then given by
\begin{eqnarray}
I_{\gamma}^{\rm E} &=& P_{\gamma \to \gamma}^{\rm MW} I_{\gamma}^{\rm MW}  +
\Bigl(1-P_{\gamma \to \gamma}^{\rm MW} \Bigr) I_{a}^{\rm MW} \ \ \ \ \ \ \ \ \ \ \ \ \ \ \ \ \ \ \ \   \nonumber  \\ 
&=& \Bigl[ \exp\left(-\tau_\gamma\right) P_{\gamma \to \gamma}^{\rm MW} P_{\gamma \to \gamma}^{\rm CL} 
+ \Bigl(1-P_{\gamma \to \gamma}^{\rm MW} \Bigr) \Bigl(1-P_{\gamma \to \gamma}^{\rm CL} \Bigr) \Bigr] I_{\gamma}^0.
\end{eqnarray}
We remind that this expression is valid only to get  the average  photon flux. In general, in order to calculate
the observable photon flux at Earth one should propagate  the full
evolution matrix inside the cluster, the intergalactic medium and the Milky Way.

Finally, we stress that the stochastic nature of the $\gamma \to a$ conversions in the magnetic field of a galaxy cluster implies that the final conversion probability can be considerably different from the average since the true magnetic field configuration along the line of sight is unknown. This fact entails that the photon flux observed at Earth should be better characterized in terms of probability distribution functions, obtained by considering $\gamma \to a$ conversions over different realizations of the intra-cluster magnetic field. An example of these distributions is shown in Fig.~\ref{fig3} for the source 1ES 0414+009 at different energies. These 
distributions have been found by simulating $10^7$ different realizations of the intra-cluster magnetic field. For simplicity we have normalized the distributions to their maxima.
 In particular, the distributions at low and high energies reflect the distribution of $I_\gamma$ and $I_a$, respectively, at the exit of the cluster. At low energies ($E\sim 100$~GeV), since the conversion probability $P_{a \to \gamma}$ in the Milky Way is small (indeed, we are close to the critial energy $E_c$), the dominant contribution comes from the emitted -- almost unabsorbed -- photon flux. Conversely, at high energies ($E>10$~TeV) almost all the emitted photons are absorbed and so the final photon flux is practically due to the reconverted ALPs in the Milky Way. At intermediate energies the distributions are a sort of ``combination'' between the emitted photon and ALP distributions. 
From Fig.~\ref{fig3} it is also obvious that the distributions are highly skewed with a changing asymmetry from low to high energies.
We also must keep in mind that since photons of different energies cross the same (unknown) configuration of magnetic field,
the distributions at different energies are not independent. The study of the correlations of the distributions at different energies is however beyond the scope of this work.

%
\begin{figure}[!h]  
\centering
\includegraphics[angle=0,width=0.75\columnwidth]{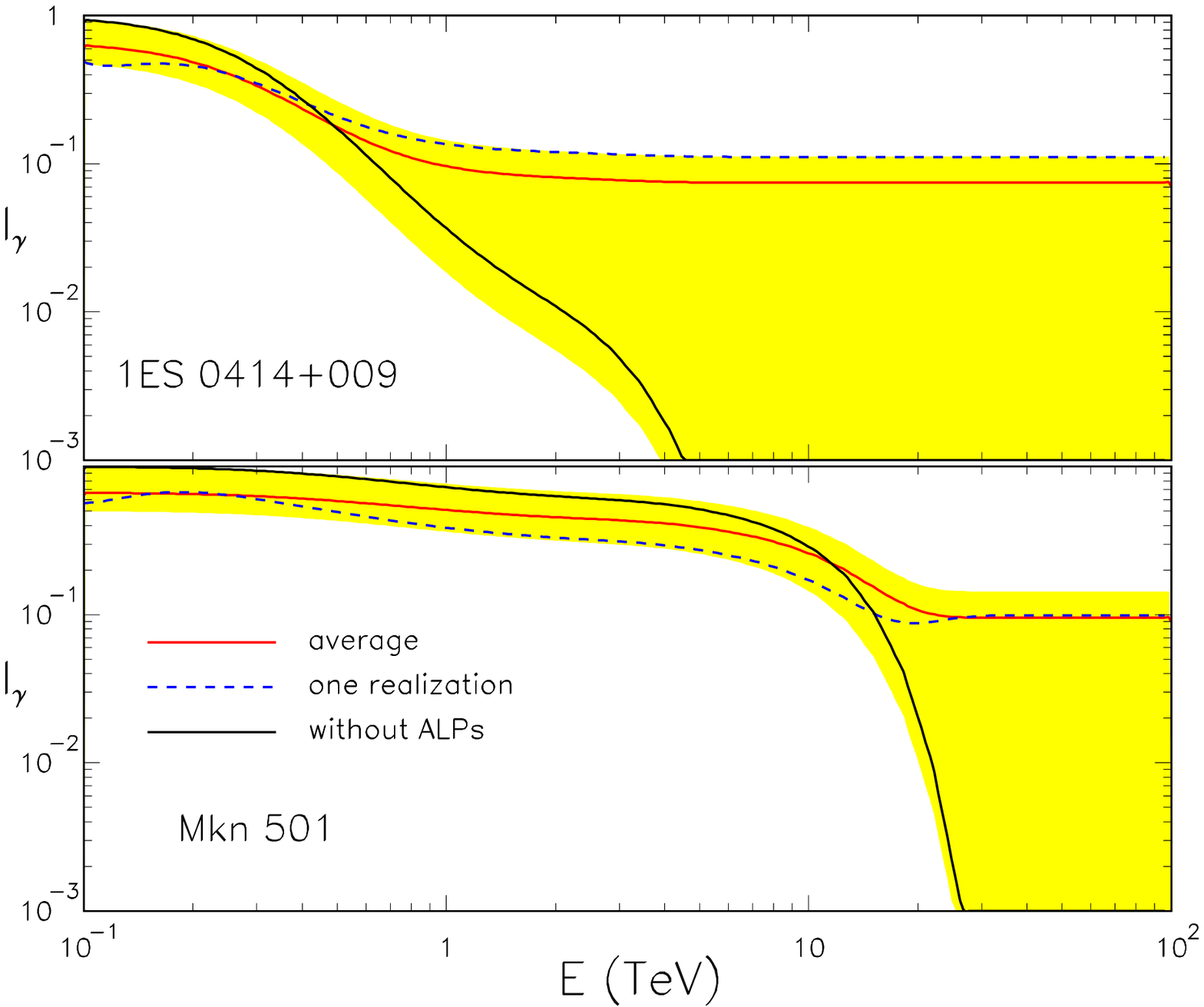} 
\includegraphics[angle=0,width=0.75\columnwidth]{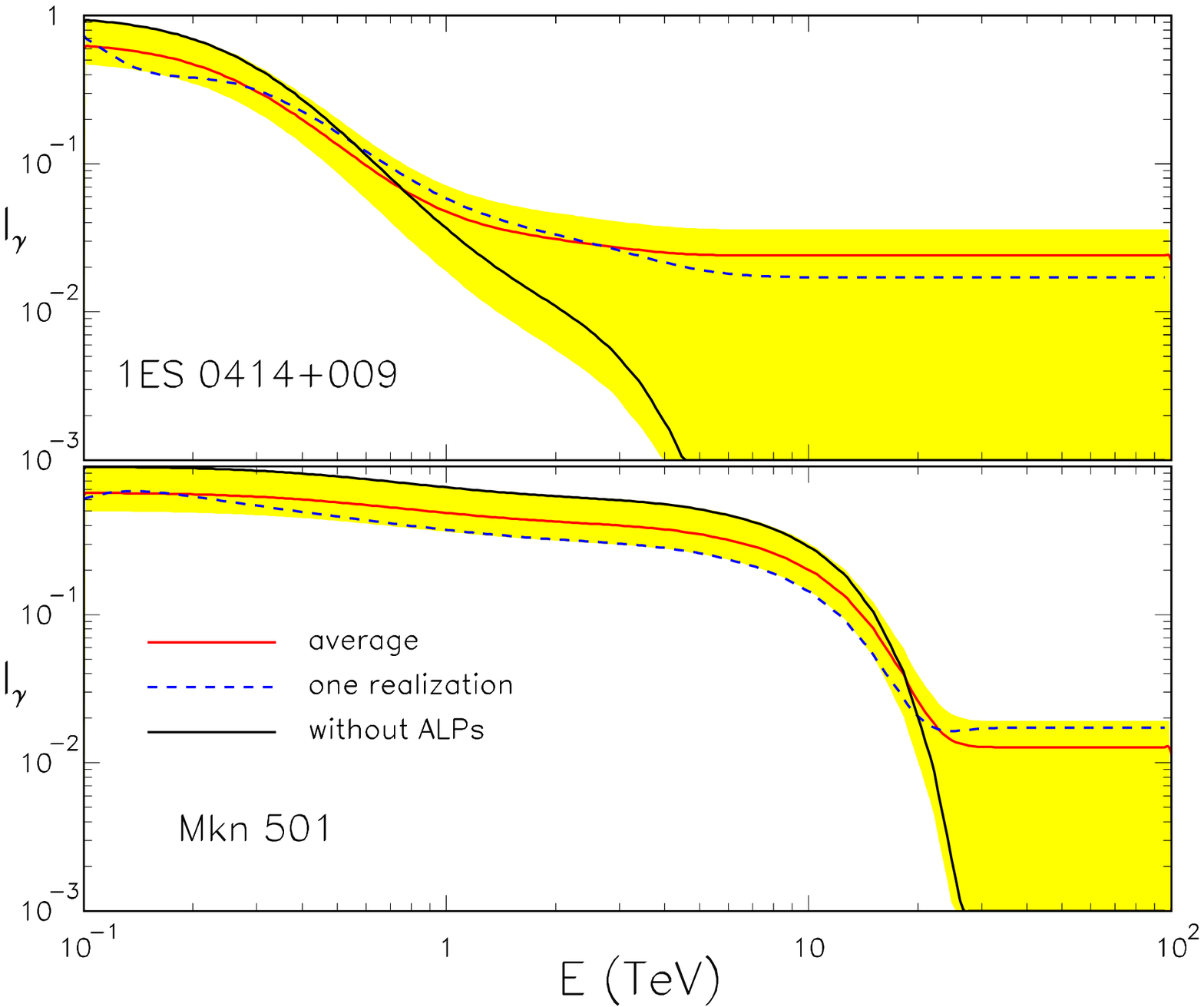} 
 \caption{Observable photon flux $I_{\gamma}$ (the emitted photon flux is normalized to 1) as a function of energy for 1ES 0414+009 and for Mkn 501  using the Jansson and Farrar model~\cite{Jansson:2012pc} (upper plot) 
 and the Pshirkov et al.  model~\cite{Pshirkov:2011um} (lower plot) respectively. The black solid curve represents the flux expected in the presence of only absorption onto EBL. We are using the FRV EBL model~\cite{Franceschini:2008tp,Franceschiniweb}. The solid red curve represents the average photon flux in the presence of $\gamma \to a \to \gamma$ conversions. The dashed curve corresponds to conversions for a particular realization of the intra-cluster magnetic field. The shaded band is the envelope of the results on all the possible realizations of the intra-cluster magnetic field.}
\label{fig4}  
\end{figure}  
%
%
%
%

\section{Results}

After having described the details of the proposed mechanism, we discuss the observational signatures concerning the energy spectra of VHE photon sources.  As  specific examples we consider again the energy spectra of 1ES0414+009 at redshift $z=0.287$ and Mkn 501 at $z=0.034$. 
In order to show the effect of $\gamma \to a \to \gamma$ conversions on VHE photons, we exhibit in Fig.~\ref{fig4} the observable photon flux $I_{\gamma}$ (normalized to the emitted one) as a function of energy for 1ES 0414+009  and for Mkn 501 for both the models of Jansson and Farrar \cite{Jansson:2012pc} (upper plot) and 
Pshirkov et al.  \cite{Pshirkov:2011um} (lower plot). The solid black curves represent the flux expected in the presence of EBL absorption only. 
According to conventional physics, it turns out that the flux gets dramatically suppressed
at high energies ($E > 1$ TeV), the farthest the source the lowest the energy.  

Including the effect of $\gamma \to a \to \gamma$ conversions we see that the photon flux at high energy gets strongly enhanced with respect to the expectation in the presence of conventional physics.
 In particular, the continuous red curves represent the  photon flux in the presence of $\gamma \to a \to \gamma$ conversions, averaged over many realizations of the intra-cluster magnetic field. Indeed, the effect is striking. Since the photon-ALP conversion probability in the strong-mixing regime in which we are working is energy-independent, the photon flux displays a plateau -- instead of a sharp drop -- at which its intensity depends on the adopted Galactic magnetic field model.  On average it turns out to be between $10 \, \%$ and $3 \, \%$ of the emitted value for 1ES 0414+009, and between $10 \, \%$ and $1 \, \%$ for Mkn 501. Therefore, the existence of $\gamma \to a \to \gamma$ conversions produces a considerable hardening of the spectrum at high enough energies, thereby making it possible to detect VHE photons in a range where no observable signal would be expected according to conventional physics.

However, we remark that due to the stochastic nature of the $\gamma \to a$ conversions in the random intra-cluster magnetic fields, the observable photon flux could show large fluctuations depending on the realization of the magnetic network crossed during the propagation. In view of these fluctuations, the systematic effect of our poor knowledge of Galactic magnetic fields plays a relatively minor role (as demonstrated in Fig.~\ref{fig4}), although detailed observations of hard spectra in the VHE regime could be sensitive to the structure of the Galactic magnetic field.

An example of a particular realization is shown by the dashed curve. In this specific case we see that the observable photon flux at high energies can be even larger than the average one. However, if one considers many realizations of the intra-cluster magnetic fields one obtains the shaded band as envelope of the results. Therefore, depending on the particular magnetic realization 
crossed by the photons, it is also possible to have cases in which the suppression of the photon flux is stronger than in the presence of conventional physics. 
Nevertheless, at high energies from Fig.~3 one infers that the cases in which $I_\gamma$ is enhanced are more probable.
In general,  $\gamma \to a \to \gamma$ conversions cannot be regarded as a universal mechanism to produce an hardening in the spectrum of VHE photons. Conversely, if a large number of AGN in clusters were observed with large spectral variations this would be a supporting indication for the presence of ALPs.


\begin{figure}[!t]  
\centering
\includegraphics[angle=0,width=0.85\columnwidth]{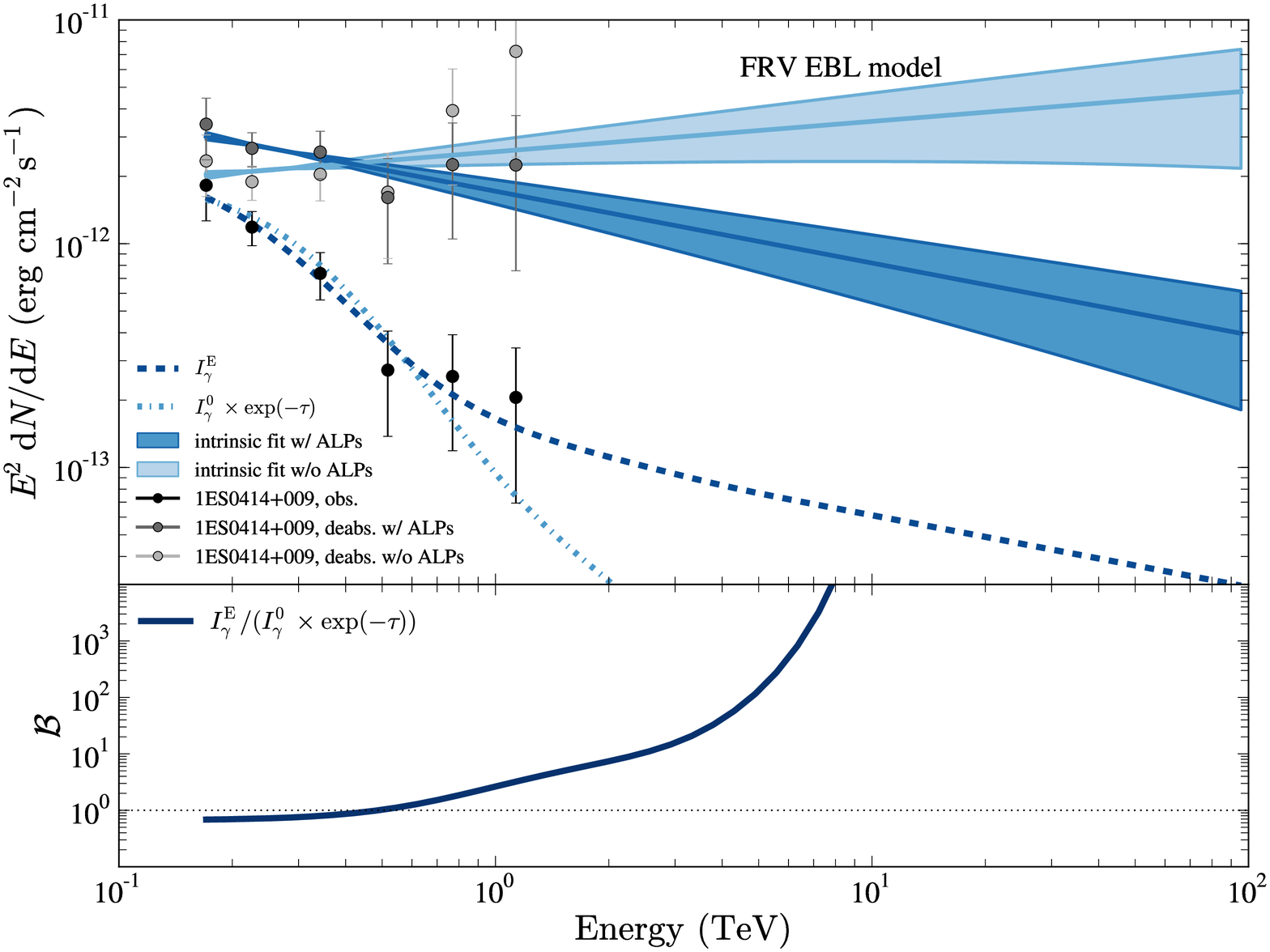} 
\includegraphics[angle=0,width=0.85\columnwidth]{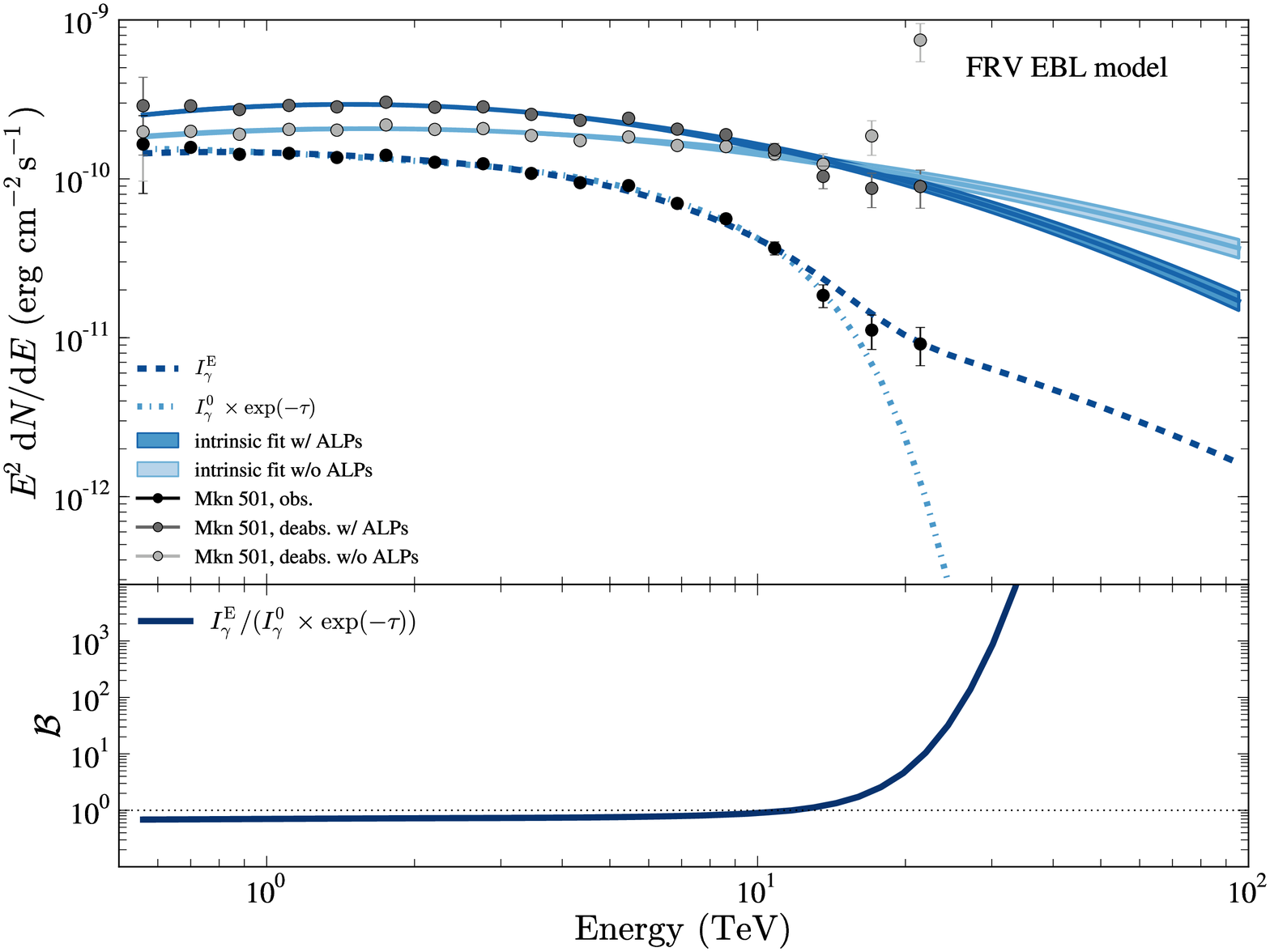} 
 \caption{
The spectra of 1ES 0414+009  \cite{HESS:2012} (upper plot) and Mkn 501 \cite{Aharonian:1999} (lower plot).
 We are using the FRV EBL model~\cite{Franceschini:2008tp,Franceschiniweb}.
 The upper panels show the observed spectra corrected for absorption with and without ALPs along with the extrapolated spectra, whereas the lower panels depict axion boost factor $\mathcal{B}$, i.e. the ratio of the observed fluxes with and without ALPs.
}

\label{fig5}  
\end{figure}

\begin{table}[t!]
 \centering
\caption{
Fit parameters for the spectra of 1ES 0414+009 and Mkn 501 corrected
with the FRV EBL model.
The spectra are described with a power law, 
$\mathrm{d}N/\mathrm{d}E = N_0 (E / E_0)^{-\Gamma}$ and a logarithmic
parabola $\mathrm{d}N/\mathrm{d}E = N_0 (E / E_0)^{-\Gamma-\beta
\ln(E/E_0)}$. The de-correlation energies are $E_0 = 0.27\,(0.28)$\,TeV
and $E_0 = 1.98$\,TeV for 1ES 0414+009 (with ALPs)  and Mkn 501 (both
scenarios), respectively, and are held fixed. 
The normalization $N_0$ is given in units of
$\mathrm{TeV}^{-1}\mathrm{cm}^{-2}\mathrm{s}^{-1}$.}
 \begin{tabular}{l|cc|cc}
  \hline
  \hline
  {} & \multicolumn{2}{c|}{1ES\,0414+009} & \multicolumn{2}{c}{Mkn
\,501}\\
  {Fit parameters} & w/o ALPs & w ALPs & w/o ALPs & w ALPs\\
  \hline
  Normalization $N_0 \times 10^{-11}$ & $2.03\pm0.24$ & $1.71\pm0.21$ &
$4.60\pm0.06$ & $3.25\pm0.04$\\
  Power-law index $\Gamma$ & $2.32\pm0.30$ & $1.86\pm0.39$ & $2.11
\pm0.02$ & $2.12\pm0.02$\\
  Curvature $\beta$ & -- & -- & $0.16\pm0.02$ & $0.10\pm0.02$\\
  $\chi^2/\mathrm{d.o.f.}$ & 0.23 & 0.60 & 1.36 & 1.94\\
  d.o.f.& 4 & 4 & 14 &  14\\
  \hline
 \end{tabular}
\label{tab:fitpar}
\end{table}


\begin{figure}[!t]  
\centering
\includegraphics[angle=0,width=0.85\columnwidth]{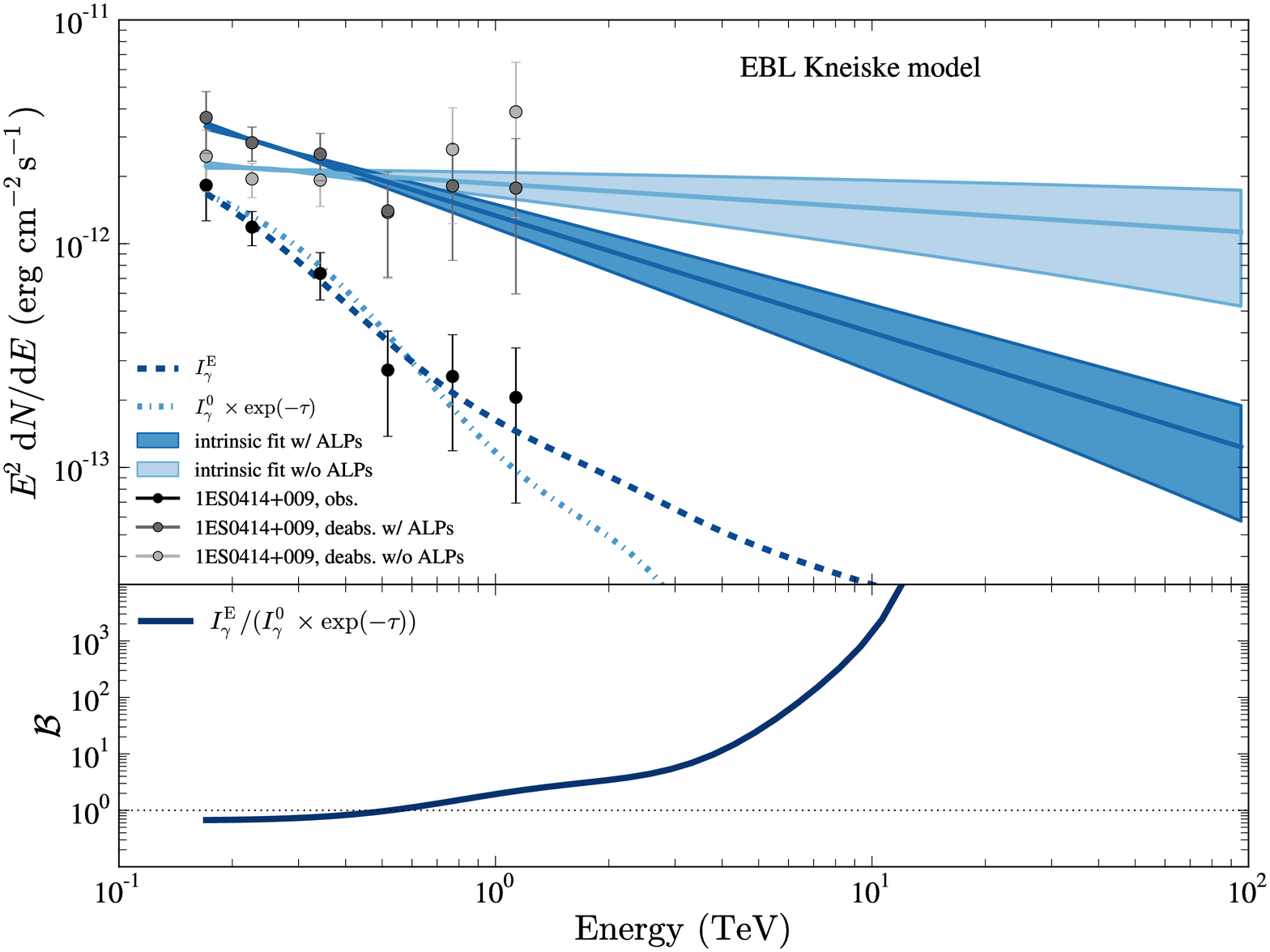} 
\includegraphics[angle=0,width=0.85\columnwidth]{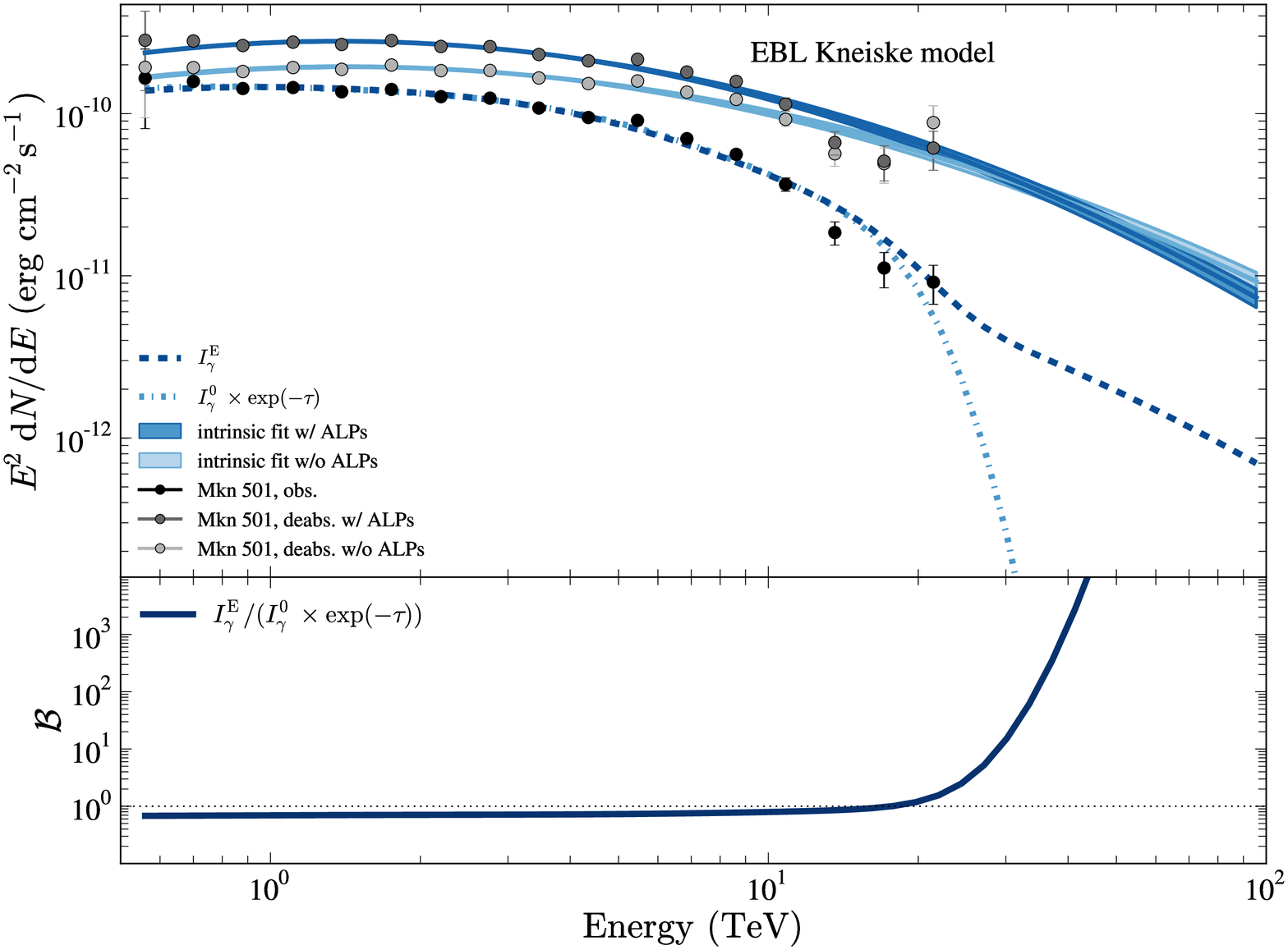} 
 \caption{
Same as Fig. \ref{fig5} but with the minimal EBL Kneiske model~\cite{Kneiske:2010,Kneiske:2010web}.}
\label{fig6}  
\end{figure}  

In Fig.~\ref{fig5} we consider the spectra of 1ES 0414+009 (upper plot) and Mkn 501 (lower plot) using the FRV EBL model~\cite{Franceschini:2008tp,Franceschiniweb}. 
The observed spectra (black bullets) are corrected for EBL absorption with and without ALPs. 
This yields the data points shown as dark and light gray bullets, respectively. 
The EBL absorption corrected spectra are subsequently fitted with a power law (1ES 0414+009) and a logarithmic parabola (Mkn 501), 
represented by the light blue (no ALPs) and dark blue bow ties (with ALPs).
The final fit parameters are shown in Table \ref{tab:fitpar}.
The difference in the observed spectra becomes prominent beyond $\sim4$\,TeV (1ES 0414+009) and $\sim30$\,TeV (Mkn 501), respectively,
where the $\gamma \to a \to \gamma$ conversion (dark blue dashed lines) predicts a substantial flux enhancement.
Unfortunately, with the present data it is  impossible to distinguish the two scenarios.
We also remark that with ALPs, the emitted photon flux at low energies 
has to be larger than in the case of conventional physics in order to describe the observations.
The reason is that more photons convert into ALPs in the cluster than the other way around in the Milky Way and the absorption at low energies is negligible.

The same analysis is repeated for the minimal EBL Kneiske model~\cite{Kneiske:2010,Kneiske:2010web} and the corresponding results are shown in Fig. \ref{fig6}.
The difference between the EBL models is marginal since the spectra are dominated at high energies by the ALPs that re-convert into photons in the Milky Way.

Finally, the boost factor of the flux intensity with respect to conventional physics
\begin{equation}
{\mathcal B}(E) = \frac{I^E_\gamma}{I^0_\gamma \exp(-\tau_\gamma)} \,\ ,
\end{equation}
is plotted in the lower panels Fig. \ref{fig5}--\ref{fig6}
for the FRV EBL model~\cite{Franceschini:2008tp,Franceschiniweb} and for the minimal EBL Kneiske model~\cite{Kneiske:2010,Kneiske:2010web}, respectively. One realizes that when ALP-photon conversions are effective, they 
can produce ${\mathcal B} \gtrsim 10^3$ thereby implying a strong enhancement of the observable photon flux with respect to expectations
based on conventional physics alone.

\section{Conclusions}

VHE $\gamma$-ray observations offer the possibility to indirectly probe the existence of ALPs predicted in many extensions of the Standard Model.
In this respect, $\gamma \to a \to \gamma$ conversions of VHE photons in the presence of cosmic magnetic fields have been recently proposed as an intriguing mechanism to explain the surprising high degree of transparency of the Universe to VHE 
photons recently observed in different high-redshift sources.
In the present work we have explored the further possibility of $\gamma \to a$ conversions in the magnetic field of galaxy clusters which frequently host VHE emitting Blazars and subsequent $a \to \gamma$ regeneration in the magnetic field of the Galaxy. We have shown that this mechanism can produce a significant hardening of the VHE photon spectrum of Blazars located in clusters of galaxies. 
More specifically, the signature of this effect in IACTs allows to infer the existence of ultra-light ALPs with mass $m_a \lesssim 10^{-8}$~eV and photon-ALP coupling $g_{a\gamma} \lesssim 10^{-10}$~GeV$^{-1}$. We expect that such a signature  would start
to emerge at energies $E\gtrsim 1$~TeV. Therefore, they are barely testable with the present generation of IACTs like H.E.S.S. \cite{Hinton:2004}, MAGIC \cite{Lorenz:2004},
 VERITAS \cite{Maier:2008}, and CANGAROO-III \cite{Kubo:2004}, covering energies in the range from $\sim$~50~GeV to $\sim$~50~TeV.
A clear-cut check can only come from the planned Cherenkov Telescope Array (CTA) \cite{Actis:2011} and High Altitude Water Cherenkov Experiment (HAWC) \cite{Sinnis:2005},
 reaching energies of 100 TeV with much higher sensitivity, or even with the HiSCORE detector, reaching PeV energies \cite{Tluczykont:2011}.

A peculiar feature of the proposed mechanism is that since the magnetic fields in galaxy clusters have a turbulent nature, the 
$\gamma \to a$ conversion probabilities present a large variance depending on the line of sight crossed by the photon/ALP beam inside the cluster. This fact suggests that our proposal could not be an universal mechanism to produce the transparency of the Universe to VHE photons. But if a large number of AGN in galaxy clusters were observed with large spectral variations this would be a positive evidence for our proposal. 

We also remark that even if an AGN is not located inside a galaxy cluster, there is a nontrivial chance that in some cases its line of sight crosses a cluster of galaxies. We plan to investigate what happens in this instance in a future work. 

Remarkably, an independent laboratory check of the ultra-light ALPs discussed in our scenario can be  performed with the planned upgrade of the photon regeneration experiment ALPS at DESY~\cite{Ehret:2010mh} and  with  the next generation solar axion detector IAXO (International Axion Observatory)~\cite{Irastorza:2011gs}.
This confirms once more the nice synergy between astrophysical and laboratory searches to find axion-like particles.

\section*{Acknowledgments} 

We thank Carmelo Evoli, Dario Grasso, and Fabrizio Tavecchio   for interesting discussions on AGN and magnetic fields in galaxy clusters. We also thank R. Jansson for providing the details of the implementation of model \cite{Jansson:2012pc}.
We thank Georg Raffelt for reading the manuscript and for comments on it.
D.M. acknowledges kind hospitality at the II Institute for Theoretical Physics at the University of Hamburg where part of this work was done.
L.M. acknowledges support from the Alexander von Humboldt foundation.
M.M. would like to thank the state excellence cluster ``Connecting Particles with the Cosmos'' at the University of Hamburg.
The work of  A.M. was supported by the German Science Foundation (DFG) within the Collaborative Research Center 676 ``Particles, Strings and the Early Universe''. 
 The work of D.M. was partly supported by the Italian MIUR and INFN through the ``Astroparticle Physics'' research project.



\end{document}